\documentclass[times,final]{elsarticle}
\usepackage{jcomp,framed,multirow,amssymb,latexsym,url,xcolor,dsfont}
\definecolor{newcolor}{rgb}{.8,.349,.1}

\usepackage{hyperref}

\journal{Journal of Computational Physics}

\begin{document}

\verso{M. Lubasch, P. Moinier, D. Jaksch}

\begin{frontmatter}

\title{Multigrid Renormalization}

\author[1]{Michael Lubasch\corref{cor}}
\cortext[cor]{Corresponding author.}
\ead{michael.lubasch@physics.ox.ac.uk}
\author[2]{Pierre Moinier}
\author[1,3,4]{Dieter Jaksch}

\address[1]{Clarendon Laboratory, University of Oxford, Parks Road, Oxford OX1 3PU, United Kingdom}
\address[2]{BAE Systems MAI, Computational Engineering, Buckingham House, FPC 267 PO Box 5, Filton, Bristol BS34 7QW, United Kingdom}
\address[3]{Centre for Quantum Technologies, National University of Singapore, 3 Science Drive 2, 117543 Singapore}
\address[4]{Keble College, University of Oxford, Parks Road, Oxford OX1 3PG, United Kingdom}

\received{\ldots}
\finalform{\ldots}
\accepted{\ldots}
\availableonline{\ldots}

\begin{abstract}
We combine the multigrid (MG) method with state-of-the-art concepts from the variational formulation of the numerical renormalization group.
The resulting MG renormalization (MGR) method is a natural generalization of the MG method for solving partial differential equations.
When the solution on a grid of $N$ points is sought, our MGR method has a computational cost scaling as $\mathcal{O}(\log(N))$, as opposed to $\mathcal{O}(N)$ for the best standard MG method.
Therefore MGR can exponentially speed up standard MG computations.
To illustrate our method, we develop a novel algorithm for the ground state computation of the nonlinear Schr\"{o}dinger equation.
Our algorithm acts variationally on tensor products and updates the tensors one after another by solving a local nonlinear optimization problem.
We compare several different methods for the nonlinear tensor update and find that the Newton method is the most efficient as well as precise.
The combination of MGR with our nonlinear ground state algorithm produces accurate results for the nonlinear Schr\"{o}dinger equation on $N = 10^{18}$ grid points in three spatial dimensions.
\end{abstract}


\end{frontmatter}

\section{Introduction}

Multigrid (MG) methods~\cite{Br77} are among the most efficient tools for the boundary value problem of (nonlinear) partial differential equations~\cite{Ha85}.
The standard MG method discretizes space by a finite number of grid points $N$ and provides a recipe for approaching the solution in continuous space by successively increasing $N$.
The ultimate goal of MG methods is the continuous solution which is obtained in the limit $N \to \infty$.
MG methods are iterative solvers and their computational cost scales better than for others, e.g.\ direct solvers, namely linearly with $N$ in the best case.
This allows MG methods to go to larger values of $N$ and closer to the continuum limit $N \to \infty$ than alternative methods.

Around the same time when MG methods were first formulated to tackle the limit $N \to \infty$ grid points in mathematics, new renormalization group (RG) techniques~\cite{Wi75} were invented to address the limit of $N \to \infty$ quantum particles in theoretical physics.
The invention of the numerical renormalization group (NRG)~\cite{Wi75} led to a breakthrough when it solved the Kondo problem -- of a single magnetic quantum impurity in a nonmagnetic metal -- which was an outstanding problem in condensed matter theory until then.
While the NRG has proven to be a very powerful tool for a variety of quantum impurity problems~\cite{BuCoPr08}, it has fundamental difficulties with more general quantum many-body problems such as e.g.\ quantum spin chains.
For the latter systems, most of these difficulties could be overcome when the density matrix renormalization group (DMRG) was invented~\cite{Wh92, Wh93} as a modified version of NRG -- reference~\cite{Sc05} summarizes the early success of DMRG.
However, also DMRG has fundamental limitations, and many of them could be overcome when yet another key discovery was made: namely that DMRG can be equally formulated as a variational algorithm over an ansatz called matrix product states~\cite{OeRo95}.
This led to new variational renormalization group (RG) methods that generalized and extended DMRG -- references~\cite{VeMuCi08, CiVe09, Sc11} cover this topic comprehensively.
Note that also NRG could be improved by reformulation as a variational RG method~\cite{WeVeScCiDe09, PiVe12}.

\begin{figure}
\centering
\includegraphics[width=57.596mm]{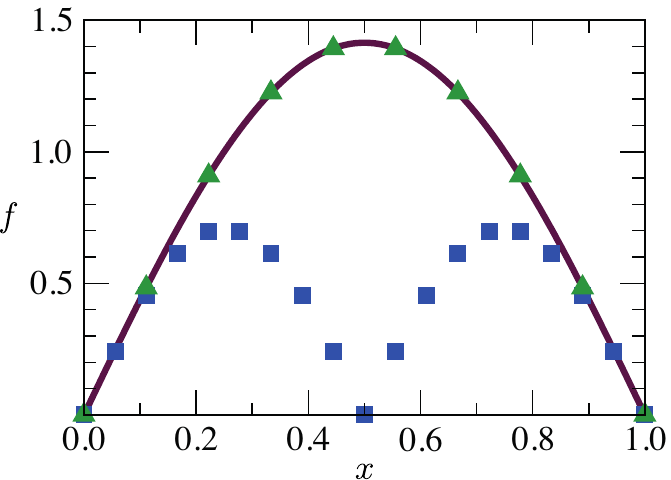}
\caption{\label{fig:1}
Exact ground state wave function $f(x)$ for a free quantum particle in a box (solid line).
The numerical renormalization group (NRG) approximates the solution on a finer grid using the solutions from a coarser grid as depicted by the squares, which correspond to the lowest eigenstate on the coarser grid.
More precisely, NRG uses many low-lying eigenstates on the coarser grid to approximate the solution on the finer grid.
The standard NRG approach fails because all combinations of these eigenstates have a node at $x = 0.5$ which is absent in the desired solution on the finer grid.
Applying a variety of boundary conditions on the coarser grid improves NRG for this problem~\cite{WhNo92}.
However, the optimal boundary conditions on the coarser grid need to be found empirically, and so this problem is a classic example for the limitations of NRG.
Our multigrid renormalization (MGR) method approximates the solution on a finer grid using the solution from a coarser grid as depicted by the triangles in a natural and optimal way.
Our MGR method uses the same boundary conditions on the coarser grid as on the finer grid, and these are, in fact, the desired boundary conditions for the continuous solution.
}
\end{figure}

In this article, we generalize and extend the MG method by combining it with state-of-the-art concepts from variational RG methods.
The resulting MG renormalization (MGR) method has a computational cost that formally scales like $\mathcal{O}\left(\log(N)\right)$ and can thus be exponentially faster than the MG method.
Our MGR method is a natural combination of MG and variational RG methods, and it is optimal in a certain sense that we will explain later.
As an example, figure~\ref{fig:1} compares the performance of NRG and our MGR method for a free quantum particle in a box.
This problem has often been used to illustrate fundamental limitations of NRG~\cite{WhNo92}.
Our MGR method overcomes these limitations.

Later in this article, we will see that MGR gives rise to tensor product states.
This variational ansatz is also known as tensor network states or as tensor networks in the physics community -- and references~\cite{VeMuCi08, CiVe09, Or14} give a comprehensive overview of the physics perspective.
An important subset of tensor product states is called matrix product states~\cite{OeRo95, FaNaWe92, Vi03, PeVeWoCi07}.
These are also known as tensor train decompositions~\cite{Os11} or as quantics tensor train decompositions~\cite{Os10, Kh11} in the mathematics community.
Here we will use existing results from the mathematics community.
However, our motivation, derivations, and applications originate from a physics perspective.

It is also interesting to note that MGR produces a function representation that was originally proposed in the context of quantum computation~\cite{Wi96, Za97, Za98}.
There, this representation was used to show that a quantum computer achieves an exponential compression rate of lossless compression compared to a classical computer.
Here, by using tensor product states, we will understand that MGR achieves an exponential compression rate of lossy compression on a classical computer.

This article is structured as follows.
In section~\ref{sec:mgr} we present the general MGR concept.
We show how MGR naturally leads to tensor product states in subsection~\ref{subsec:tps} and formulate the MGR method for the Poisson equation in subsection~\ref{subsec:pois}.
MGR is then used to construct a new algorithm for the ground state computation of the nonlinear Schr\"{o}dinger equation in section~\ref{sec:nlse}.
There we first introduce quantities that measure the accuracy of MGR in subsection~\ref{subsec:prec}, then present our new algorithm in subsection~\ref{subsec:algo}, and finally analyse its performance in subsection~\ref{subsec:perf}.

\section{Multigrid renormalization (MGR) method}\label{sec:mgr}

In this section we present the MGR method, which is a general concept that can be applied to any (nonlinear) partial differential equation.

\subsection{MGR gives rise to tensor product states}\label{subsec:tps}

In the following we use the finite difference method to represent (nonlinear) partial differential equations.
We denote the sought solution function by $f$ and assume that it is defined on the interval $[0, 1]$ in each of its variables -- any function can be scaled to satisfy this with all its variables.
We define the vector $|f\rangle_{h}$ as the values of $f$ on an equidistant grid of spacing $h$.
Furthermore each variable of $f$ shall be discretized by $N = 2^{L}$ points, i.e.\ $h = 1/N = 1/2^{L}$ for each variable.
If $L$, $N$, or $h$ is given explicitly then we drop the subindex $h$ in $|f\rangle_{h}$ and write $|f\rangle$ to simplify the notation.
We first present our method for a function $f$ that depends only on one variable $x$ and later discuss the extension to more variables.

MG methods find the solution on a fine grid with the help of exact solutions from coarse grids.
A crucial ingredient is the prolongation operator $\mathcal{P}$ that maps a function from a coarse grid of spacing $h$ to a finer grid of spacing $h/2$: $\mathcal{P} |f\rangle_{h} = |f\rangle_{h/2}$.
Usually MG methods also require a restriction operator that maps a function from a fine grid to a coarser grid~\cite{Br77, Ha85, HaTr82} but we will not need this restriction operator here.

We now formulate an exact MG method that starts from a large grid spacing $h = 0.5$ and then refines the grid in an exact way.
Initially, we want to store the function values of $f$ at $x = 0.0$ and $0.5$ in a vector $|f\rangle_{h=0.5}$.
This vector lives in a two-dimensional vector space that is spanned by two basis vectors $|0\rangle$ and $|1\rangle$.
We define that the basis vector $|l_{1}\rangle$ -- where $l_{1} = 0$ or $1$ -- corresponds to $x = l_{1}2^{-1}$ on the grid.
We write $|f\rangle_{h=0.5} = \sum_{l_{1} = 0}^{1} F[1]^{l_{1}} |l_{1}\rangle$ where the tensor $F[1]^{l_{1}}$ has one index $l_{1}$ that takes two values $0$ or $1$ for the two components of $|f\rangle_{h=0.5}$ with respect to the basis vectors $|0\rangle$ and $|1\rangle$.
We define a new tensor $F[1]_{\alpha_{1}}^{l_{1}}$ where $l_{1}$ is the same as before and $\alpha_{1}$ takes two values, and we write any new two basis vectors $|\alpha_{1}\rangle = \sum_{l_{1} = 0}^{1} F[1]_{\alpha_{1}}^{l_{1}} |l_{1}\rangle$ in terms of the old basis vectors $|0\rangle$ and $|1\rangle$.
Now we decrease the grid spacing by a factor $0.5$ such that $h = 0.25$.
For this refined grid, we want to store the function values of $f$ at $x = 0.0$, $0.25$, $0.5$, and $0.75$ in a vector $|f\rangle_{h=0.25}$.
This vector lives in a four-dimensional vector space that we construct as the tensor product of the previous vector space with basis vectors $|\alpha_{1}\rangle$ and a new two-dimensional vector space with basis vectors $|0\rangle$ and $|1\rangle$.
We define that the basis vector $|l_{1}\rangle |l_{2}\rangle$ -- where $l_{1} = 0$ or $1$ and $l_{2} = 0$ or $1$ -- corresponds to $x = l_{1}2^{-1} + l_{2}2^{-2}$ on the grid.
We write $|f\rangle_{h=0.25} = \sum_{l_{2}} \sum_{\alpha_{1}} F[2]_{\alpha_{1}}^{l_{2}} |\alpha_{1}\rangle |l_{2}\rangle$ where the tensor $F[2]_{\alpha_{1}}^{l_{2}}$ has one index $\alpha_{1}$ for the previous basis and one index $l_{2}$ for the new basis.
We define a new tensor $F[2]_{\alpha_{1}, \alpha_{2}}^{l_{2}}$ where $\alpha_{1}$ and $l_{2}$ are the same as before and $\alpha_{2}$ takes four values, and we write any new four basis vectors $|\alpha_{2}\rangle = \sum_{l_{2}} \sum_{\alpha_{1}} F[2]_{\alpha_{1}, \alpha_{2}}^{l_{2}} |\alpha_{1}\rangle |l_{2}\rangle$ in terms of the old basis vectors.
We continue this procedure.
In each step the grid spacing $h$ decreases by a factor of $0.5$ and the dimensionality of the corresponding vector $|f\rangle$ grows by a factor of $2$.
Thus in step $\ell$, the grid spacing reads $h = 1/2^{\ell}$ and $|f\rangle_{h = 1/2^{\ell}}$ lives in a $2^{\ell}$-dimensional vector space.
We define that the basis vector $|l_{1}\rangle |l_{2}\rangle \ldots |l_{\ell}\rangle$ -- where $l_{1} = 0$ or $1$, $l_{2} = 0$ or $1$, and so on -- corresponds to $x = l_{1}2^{-1} + l_{2}2^{-2} + \ldots + l_{\ell}2^{-\ell}$ on the grid.
We write $|f\rangle_{h = 1/2^{\ell}} = \sum_{l_{\ell}} \sum_{\alpha_{\ell-1}} F[\ell]_{\alpha_{\ell-1}}^{l_{\ell}} |\alpha_{\ell-1}\rangle |l_{\ell}\rangle$.
And we define a new tensor $F[\ell]_{\alpha_{\ell-1}, \alpha_{\ell}}^{l_{\ell}}$ to write any new $2^{\ell}$ basis vectors $|\alpha_{\ell}\rangle = \sum_{l_{\ell}} \sum_{\alpha_{\ell-1}} F[\ell]_{\alpha_{\ell-1}, \alpha_{\ell}}^{l_{\ell}} |\alpha_{\ell-1}\rangle |l_{\ell}\rangle$.

If we want to store all the tensors $F[1]$, $F[2]$, \ldots on our computer, we need to be aware that in step $\ell$ the tensor $F[\ell]$ has the size $4^{\ell}$, and so the required memory grows exponentially with $\ell$.
To prevent this from happening we now define a maximum dimensionality $\chi$ for the vector space spanned by the new basis vectors $|\alpha_{\ell}\rangle$.
The $\chi$ is introduced to reduce the exponential scaling to a polynomial scaling.
For the initial steps $\ell \leq \log_{2}(\chi)$ there are $2^{\ell}$ new basis vectors, but for the later steps $\ell > \log_{2}(\chi)$ there are just $\chi$ new basis vectors.
Let us denote the dimensionality of the vector space spanned by $|0\rangle$ and $|1\rangle$ by $d$, where $d = 2$ here.
Then each tensor $F[\ell]$ has at most the size $d\chi^{2}$.
Therefore if we stop our procedure at step $L$ -- when $N = 2^{L}$ -- our memory requirements are upper-bounded by $\mathcal{O}(Ld\chi^{2})$.
We refer to our new MG procedure as a MG renormalization (MGR) method because it allows us to go to much larger values of $L$ by using renormalized degrees of freedom, given by the at most $\chi$-dimensional vector spaces.

\begin{figure}
\centering
\includegraphics[width=69.398mm]{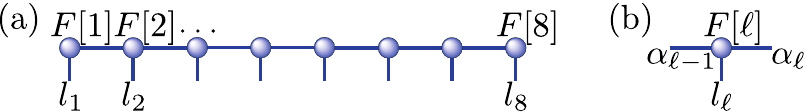}
\caption{\label{fig:2}
Matrix product state $\sum_{\alpha_{1}, \alpha_{2}, \ldots, \alpha_{7}} \left( F[1]_{\alpha_{1}}^{l_{1}} F[2]_{\alpha_{1}, \alpha_{2}}^{l_{2}} F[3]_{\alpha_{2}, \alpha_{3}}^{l_{3}} \ldots F[8]_{\alpha_{7}}^{l_{8}} \right)$ of length $L = 8$ (a) and a tensor $F[\ell]_{\alpha_{\ell-1}, \alpha_{\ell}}^{l_{\ell}}$ from an interior site $1 < \ell < L$ (b).
In this graphical representation of a tensor product state, each node (blue sphere) corresponds to a tensor and each edge (blue line) corresponds to an index.
Therefore the rank of a tensor, i.e.\ its total number of indices, is given by the total number of edges connected to that tensor.
When a tensor has an open edge then this represents a degree of freedom.
We denote the corresponding open indices by Latin letters and define that each open index takes values from $\{0, 1, \ldots, d-1\}$.
E.g.\ in (a) each open index $l_{1}$, $l_{2}$, \ldots, or $l_{L}$ can take any value from $\{0, 1, \ldots, d-1\}$ and so this matrix product state represents $N = d^{L}$ numbers.
Note that $d$ can be seen as the dimensionality of a vector space in which one open index lives.
When two tensors are connected via an edge then the corresponding two indices are summed over.
We denote these connecting indices by Greek letters and define that each connecting index takes values from $\{1, 2, \ldots, \chi\}$.
E.g.\ in (a) the tensors $F[1]$ and $F[2]$ are connected via one edge which corresponds to the sum over the index $\alpha_{1}$, i.e.\ $\sum_{\alpha_{1}} \left( F[1]_{\alpha_{1}}^{l_{1}} F[2]_{\alpha_{1}, \alpha_{2}}^{l_{2}} \right)$.
In practice the connecting indices can live in vector spaces of different dimensionalities and then $\chi$ is defined as the largest of these dimensionalities.
}
\end{figure}

Importantly, we can choose these $\chi$-dimensional vector spaces in an optimal way by making use of the fact that MGR produces tensor product states.
If we stop our MGR method at step $L$ we can write out the vector $|f\rangle_{h = 1/2^{L}} = \sum_{l_{L}} \sum_{\alpha_{L-1}} F[L]_{\alpha_{L-1}}^{l_{L}} |\alpha_{L-1}\rangle |l_{L}\rangle$ as
\begin{eqnarray}\label{eq:mps}
|f\rangle_{h = 1/2^{L}} & = & \sum_{l_{1}, l_{2}, \ldots, l_{L}} \sum_{\alpha_{1}, \alpha_{2}, \ldots, \alpha_{L-1}} \left( F[1]_{\alpha_{1}}^{l_{1}} F[2]_{\alpha_{1}, \alpha_{2}}^{l_{2}} F[3]_{\alpha_{2}, \alpha_{3}}^{l_{3}}  \ldots F[L]_{\alpha_{L-1}}^{l_{L}} \right) |l_{1}\rangle |l_{2}\rangle \ldots |l_{L}\rangle .
\end{eqnarray}
This is a matrix product state~\cite{OeRo95, FaNaWe92, Vi03, PeVeWoCi07} -- also known as tensor train decomposition~\cite{Os11} or as quantics tensor train decomposition~\cite{Os10, Kh11} -- of bond dimension $\chi$.
In a tensor product state, the term bond dimension denotes the maximum dimensionality of all the internal indices that connect different tensors.
Figure~\ref{fig:2} shows a graphical representation.
Because MGR produces tensor product states we can reformulate finite difference approaches to (nonlinear) partial differential equations as variational algorithms over tensor product states.
This allows us to make use of the plethora of already existing variational renormalization group techniques that determine the $\chi$-dimensional vector spaces in an optimal way -- see e.g.\ reference~\cite{VeMuCi08} and references therein.

Our MGR method uses tensor product states as a variational ansatz for the function $f$ and approaches the continuous solution via iteration of two steps: firstly minimizing a cost function and secondly prolonging to the next finer grid.
The cost function is a problem-specific function $c(f)$.
Here, we always choose the cost function for a problem in such a way that the desired solution $f$ minimizes the cost function.
An example cost function is the ground state energy $\langle f|H|f\rangle/\langle f|f\rangle$ where $H$ denotes a Hamiltonian (e.g.\ from finite difference discretization of a linear Schr\"{o}dinger equation).
This cost function often occurs in physical problems (particularly in quantum physics) and is minimal for the exact ground state.
If we are using a variational ansatz to approximate the exact ground state we can minimize the ground state energy via the variational parameters of that ansatz.
Using tensor product states we minimize the cost function in the spirit of alternating least squares as proposed in reference~\cite{VePoCi04}: We move from one tensor to the next and for each tensor minimize the cost function via the variational parameters of that tensor only, i.e.\ leaving the variational parameters of all other tensors fixed.
E.g., if we are using the matrix product state of equation~\ref{eq:mps} as a variational ansatz to minimize a cost function, then we start at the tensor $F[1]$ and minimize the cost function using only the variational parameters of $F[1]$.
Then we move to $F[2]$ and minimize the cost function using only the variational parameters of $F[2]$, then to $F[3]$, and so on, until $F[L]$, which is when we move backwards, i.e.\ $F[L-1]$, then $F[L-2]$, and so on, until $F[1]$.
We repeat this procedure several times until our cost function does not change significantly anymore.
In the context of tensor product states, several minimization procedures have been proposed, see e.g.\ references~\cite{SaVi07, ScWoVeCi08b} for methods using Monte Carlo sampling of the gradient.
However, to the best of our knowledge, the alternating minimization procedure of reference~\cite{VePoCi04} is the most efficient and accurate approach for the minimization of many different cost functions.
And so we use this method for the minimization of all our cost functions in this article.

\begin{figure}
\centering
\includegraphics[width=73.858mm]{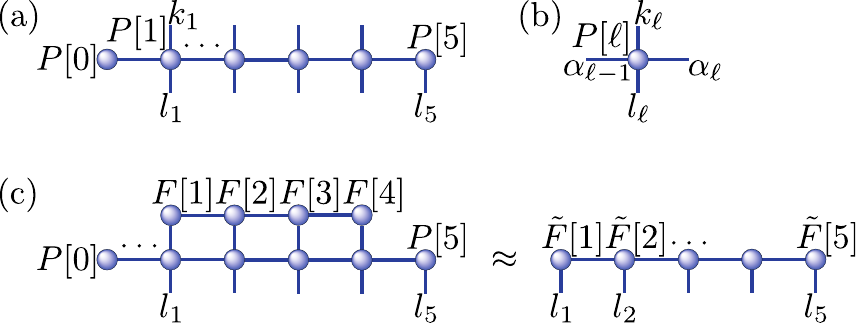}
\caption{\label{fig:3}
(a) Matrix product operator~\cite{VeGaCi04} $\sum_{\alpha_{0}, \alpha_{1}, \ldots, \alpha_{4}} \left( P[0]_{\alpha_{0}} P[1]_{\alpha_{0}, \alpha_{1}}^{k_{1}, l_{1}} \ldots P[5]_{\alpha_{4}}^{l_{5}} \right)$ for prolongation from $L = 4$ to $L+1 = 5$.
(b) A tensor $P[\ell]_{\alpha_{\ell-1}, \alpha_{\ell}}^{k_{\ell}, l_{\ell}}$ from an interior site $0 < \ell < L+1$.
(c) We prolong $|f\rangle = F[1] F[2] F[3] F[4]$ to $|\tilde{f}\rangle = \tilde{F}[1] \tilde{F}[2] \tilde{F}[3] \tilde{F}[4] \tilde{F}[5]$ by minimizing the cost function $c(|\tilde{f}\rangle) = ||\mathcal{P}|f\rangle - |\tilde{f}\rangle||^{2}$ via the alternating minimization procedure of reference~\cite{VeGaCi04}.
}
\end{figure}

After minimizing the cost function of our problem, we prolong $f$ to the next finer grid.
We can define our prolongation operator in many different ways.
Here we define the prolongation $\mathcal{P}|f\rangle_{h} = |f\rangle_{h/2}$ in such a way that the new vector $|f\rangle_{h/2}$ has the same values as $|f\rangle_{h}$ on the same grid points and on the new grid points linearly interpolates between the values of $|f\rangle_{h}$ on the two adjacent grid points.
Expressed in terms of vector components, the prolongation operator $\mathcal{P}$ maps the vector $|f\rangle_{h}$ with components $|f\rangle_{h}^{l}$ to the vector $|f\rangle_{h/2}$ with components $|f\rangle_{h/2}^{2l} = |f\rangle_{h}^{l}$ and $|f\rangle_{h/2}^{2l+1} = \left( |f\rangle_{h}^{l}+|f\rangle_{h}^{l+1} \right) / 2$.
In the tensor product formalism, the operator $\mathcal{P}$ is a tensor product operator, i.e.\ the generalization of the concept of tensor product states to operators.
In the context of matrix product states, $\mathcal{P}$ is a matrix product operator~\cite{VeGaCi04} of the form shown in figure~\ref{fig:3}.
$\mathcal{P}$ prolongs from $L$ to $L+1$ and is represented by a matrix product operator $\sum_{\alpha_{0}, \alpha_{1}, \ldots, \alpha_{L}} \left( P[0]_{\alpha_{0}} P[1]_{\alpha_{0}, \alpha_{1}}^{k_{1}, l_{1}} P[2]_{\alpha_{1}, \alpha_{2}}^{k_{2}, l_{2}} \ldots P[L]_{\alpha_{L-1}, \alpha_{L}}^{k_{L}, l_{L}} P[L+1]_{\alpha_{L}}^{l_{L+1}} \right)$.
The tensors have the entries $P[0]_{1} = 1$, $P[\ell]_{1, 1}^{0, 0} = 1 = P[\ell]_{1, 1}^{1, 1} = P[\ell]_{1, 2}^{1, 0} = P[\ell]_{2, 2}^{0, 1}$ for $0 < \ell < L+1$, $P[L+1]_{1}^{0} = 1$ and $P[L+1]_{1}^{1} = 0.5 = P[L+1]_{2}^{1}$, and all other tensor entries are zero.
Remember that according to our notational convention the $\alpha$ indices run from $1$ to at most $\chi$ whereas the $k$ and $l$ indices take values from $0$ to $d-1$, as explained in figure~\ref{fig:2}.
We see that our matrix product operator for $\mathcal{P}$ has bond dimension $2$.
We prolong $|f\rangle_{h}$ to $|\tilde{f}\rangle_{h/2}$ approximately, by minimizing the cost function $c(|\tilde{f}\rangle_{h/2}) = ||\mathcal{P}|f\rangle_{h} - |\tilde{f}\rangle_{h/2}||^{2}$ via the alternating minimization procedure of reference~\cite{VeGaCi04}.
This approximate prolongation has a computational cost that scales like $\mathcal{O}(Ld\chi^{3}) + \mathcal{O}(Ld^{2}\chi^{2})$.
An exact prolongation is not computationally efficient.
If we were to compute $\mathcal{P}|f\rangle_{h} = |f\rangle_{h/2}$ exactly, then the bond dimension $\chi$ of $|f\rangle_{h/2}$ would be a factor of $2$ larger than the bond dimension of $|f\rangle_{h}$.
Thus, for an exact prolongation, the bond dimension doubles after each prolongation, and so $\chi$ grows exponentially with the number of prolongations.
Remarkably, for all cases in this article, the approximation errors of our approximate prolongations turned out to be negligibly small, as we have checked explicitly.

In our MGR method, we can choose among many different tensor product ansatzes for higher spatial dimensions $\mathrm{dim} > 1$.
In this article, we simply use matrix product states with $d = 2^{\mathrm{dim}}$ as our ansatz for all higher dimensions.
Each index $l_{\ell}$ now represents a multi-index $l_{\ell} = (l_{\ell}^{\mathrm{x}}, l_{\ell}^{\mathrm{y}})$ in two or $l_{\ell} = (l_{\ell}^{\mathrm{x}}, l_{\ell}^{\mathrm{y}}, l_{\ell}^{\mathrm{z}})$ in three spatial dimensions.
And the individual indices in this multi-index specify the spatial coordinates $(x, y)$ in two or $(x, y, z)$ in three spatial dimensions as  $x = \sum_{\ell=1}^{L} l_{\ell}^{\mathrm{x}} 2^{-l_{\ell}^{\mathrm{x}}}$, $y = \sum_{\ell=1}^{L} l_{\ell}^{\mathrm{y}} 2^{-l_{\ell}^{\mathrm{y}}}$, and $z = \sum_{\ell=1}^{L} l_{\ell}^{\mathrm{z}} 2^{-l_{\ell}^{\mathrm{z}}}$.
For the mapping from the multi-index to the single index, one usually chooses either lexicographic order -- where the last index runs fastest (which corresponds to row-major order for a two-dimensional array) -- or colexicographic order -- where the first index runs fastest (which corresponds to column-major order for a two-dimensional array) -- and in our case it is not important which order one chooses.
The prolongation operators for higher spatial dimensions are obtained from the matrix product operator $\mathcal{P}$ for one spatial dimension.
We obtain the matrix product operator $\mathcal{P} \otimes \mathcal{P}$ of bond dimension $4$ in two spatial dimensions and $\mathcal{P} \otimes \mathcal{P} \otimes \mathcal{P}$ of bond dimension $8$ in three spatial dimensions, where $\otimes$ denotes the Kronecker product.

\subsection{MGR for the Poisson equation}\label{subsec:pois}

The Poisson equation is a paradigmatic application for MG methods.
After discretization, this equation reads $\Delta_{h} |f\rangle_{h} = |g\rangle_{h}$ where $\Delta_{h}$ represents the Laplace operator, $|g\rangle_{h}$ and boundary conditions are given, and $|f\rangle_{h}$ is sought.
We define the Laplace operator via the usual finite difference approximation of the second derivative from a Taylor expansion.
Expressed in terms of vector components, our Laplace operator $\Delta_{h}$ maps the vector $|f\rangle_{h}$ with components $|f\rangle_{h}^{l}$ to the vector $\Delta_{h} |f\rangle_{h}$ with components $(\Delta_{h} |f\rangle_{h})^{l} = \left( |f\rangle_{h}^{l+1}-2|f\rangle_{h}^{l}+|f\rangle_{h}^{l-1} \right) / h^{2}$.
This finite difference approximation of the second derivative $\Delta_{h}$ has an error $\mathcal{O}(h^{2})$ that we can reduce systematically by decreasing $h$, i.e.\ increasing $L$.
We choose the function $g$ on the right-hand side as a polynomial because then we know the exact solution $f$ of our Poisson equation.
This allows us to analyze the performance of our MGR method by comparing to the exact solution.
We choose the boundary conditions $f(x = -h) = 0 = f(x = 1)$ -- and similarly in higher dimensions for $y$ and $z$ -- as they can be implemented easily in our representation of the Laplace operator.

\begin{figure}
\centering
\includegraphics[width=71.142mm]{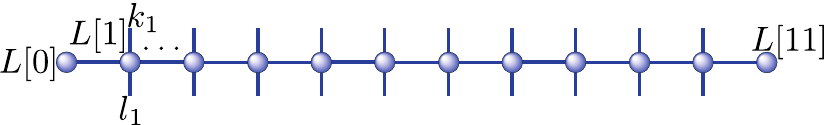}
\caption{\label{fig:4}
Laplace operator after discretization for $L = 10$ -- i.e.\ for $d^{10}$ grid points where $d = 2^{\mathrm{dim}}$ -- represented by a matrix product operator $\sum_{\alpha_{0}, \alpha_{1}, \ldots, \alpha_{10}} L[0]_{\alpha_{0}} L[1]_{\alpha_{0}, \alpha_{1}}^{k_{1}, l_{1}} L[2]_{\alpha_{1}, \alpha_{2}}^{k_{2}, l_{2}} \ldots L[10]_{\alpha_{9}, \alpha_{10}}^{k_{10}, l_{10}} L[11]_{\alpha_{10}}$.
The entries of each tensor $L[\ell]$ for $0 \leq \ell \leq L+1$ are defined and explained in the text.
}
\end{figure}

The Laplace operator is written as a matrix product operator as shown in figure~\ref{fig:4}, see also references~\cite{Os11, Os10, KaKh12}.
In one spatial dimension this reads $\sum_{\alpha_{0}, \alpha_{1}, \ldots, \alpha_{L}} L[0]_{\alpha_{0}} L[1]_{\alpha_{0}, \alpha_{1}}^{k_{1}, l_{1}} L[2]_{\alpha_{1}, \alpha_{2}}^{k_{2}, l_{2}} \ldots L[L]_{\alpha_{L-1}, \alpha_{L}}^{k_{L}, l_{L}} L[L+1]_{\alpha_{L}}$.
The tensors have the entries $L[0]_{1} = 1$, $L[\ell]_{1, 1}^{0, 0} = 4 = L[\ell]_{1, 1}^{1, 1} = L[\ell]_{2, 2}^{1, 0} = L[\ell]_{1, 2}^{0, 1} = L[\ell]_{1, 3}^{1, 0} = L[\ell]_{3, 3}^{0, 1}$ for $0 < \ell < L+1$, $L[L+1]_{1} = -2$ and $L[L+1]_{2} = 1 = L[L+1]_{3}$, and all other tensor entries are zero.
This matrix product operator has bond dimension $3$.
Our Laplace operators for higher spatial dimensions are constructed from this matrix product operator $\Delta_{h}$ for one spatial dimension.
We obtain the matrix product operator $\Delta_{h} \otimes \mathds{1} + \mathds{1} \otimes \Delta_{h}$ of bond dimension $6$ in two spatial dimensions and $\Delta_{h} \otimes \mathds{1} \otimes \mathds{1} + \mathds{1} \otimes \Delta_{h} \otimes \mathds{1} + \mathds{1} \otimes \mathds{1} \otimes \Delta_{h}$ of bond dimension $9$ in three spatial dimensions, where $\mathds{1}$ denotes the identity operator for one spatial dimension.

\begin{figure}
\centering
\includegraphics[width=74.037mm]{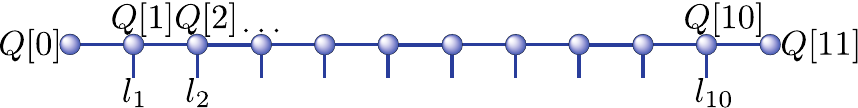}
\caption{\label{fig:5}
Polynomial function after discretization for $L = 10$ -- i.e.\ for $d^{10}$ grid points where $d = 2^{\mathrm{dim}}$ -- represented by a matrix product state $\sum_{\alpha_{0}, \alpha_{1}, \ldots, \alpha_{10}} Q[0]_{\alpha_{0}} Q[1]_{\alpha_{0}, \alpha_{1}}^{l_{1}} Q[2]_{\alpha_{1}, \alpha_{2}}^{l_{2}} \ldots Q[10]_{\alpha_{9}, \alpha_{10}}^{l_{10}} Q[11]_{\alpha_{10}}$.
The entries of each tensor $Q[\ell]$ for $0 \leq \ell \leq L+1$ are defined and explained in the text.
}
\end{figure}

Polynomial functions are written as matrix product states as shown in figure~\ref{fig:5}, see also references~\cite{Kh11, Os13}.
We consider a polynomial $q(x) = \sum_{l=0}^{\kappa} c_{l} x^{l}$ of degree $\kappa$.
After discretization, the matrix product state
\begin{eqnarray*}
\sum_{\alpha_{0}, \alpha_{1}, \ldots, \alpha_{L}} Q[0]_{\alpha_{0}} Q[1]_{\alpha_{0}, \alpha_{1}}^{l_{1}} Q[2]_{\alpha_{1}, \alpha_{2}}^{l_{2}} \ldots Q[L]_{\alpha_{L-1}, \alpha_{L}}^{l_{L}} Q[L+1]_{\alpha_{L}}
\end{eqnarray*}
for such a polynomial has bond dimension $\kappa+1$.
The tensors have the entries $Q[0]_{1} = 1$, $Q[\ell]_{\alpha, \alpha}^{0} = 1 = Q[\ell]_{\alpha, \alpha}^{1} \forall \alpha \in \{1, 2, \ldots, \kappa+1\}$ and $Q[\ell]_{\alpha, \beta}^{1} = {\beta-1 \choose \alpha-1} \cdot 2^{-(\beta-\alpha)\ell} \forall \alpha < \beta: \beta \in \{2, 3, \ldots, \kappa+1\}$ for $0 < \ell < L+1$, $L[L+1]_{\alpha} = c_{\alpha-1} \forall \alpha \in \{1, 2, \ldots, \kappa+1\}$, and all other tensor entries are zero.
Here, we construct higher-dimensional polynomial functions from the matrix product state $Q_{h}$ for one spatial dimension.
We use $Q_{h}^{\mathrm{x}} \otimes Q_{h}^{\mathrm{y}}$ in two and $Q_{h}^{\mathrm{x}} \otimes Q_{h}^{\mathrm{y}} \otimes Q_{h}^{\mathrm{z}}$ in three spatial dimensions.
Therefore our higher-dimensional polynomials are separable products of one-dimensional polynomials, which suffice for this analysis here.

Here we compare the two cost functions $c_{1}(|f\rangle) = ||\Delta |f\rangle - |g\rangle||^{2}$ and $c_{2}(|f\rangle) = \langle f| \Delta |f\rangle - 2\langle f|g\rangle$.
Note that the first cost function can be written out as $c_{1}(|f\rangle) = \langle f| \Delta^{T} \Delta |f\rangle - \langle f| \Delta^{T} |g\rangle - \langle g| \Delta |f\rangle + \langle g|g\rangle$ -- where the superscript $T$ denotes transposition -- and is minimal and zero for the solution of the Poisson equation.
The second cost function is minimal and possibly non-zero (having the value $-\langle f|g\rangle$) for the solution of the Poisson equation.
This statement is true because the gradient of $c_{2}$ with respect to $|f\rangle$ is $2\Delta|f\rangle - 2|g\rangle$ and vanishes at the minimum $|f^{\mathrm{min}}\rangle$ of $c_{2}$ such that $2\Delta|f^{\mathrm{min}}\rangle - 2|g\rangle = 0$ which is the desired solution of the Poisson equation $\Delta|f^{\mathrm{min}}\rangle = |g\rangle$.
We minimize these cost functions using the alternating minimization procedure of reference~\cite{VePoCi04}.
During this procedure, the tensors of our tensor product state are updated one after another and in each tensor update the cost function is minimized variationally using only the parameters of the tensor that is updated while the parameters of all the other tensors are fixed.
For each tensor update, we compare the pseudoinverse, steepest descent, and conjugate gradient.
The pseudoinverse has the computational cost $\mathcal{O}(d^{3}\chi^{6})$ and finds the optimal tensor parameters in one step.
Steepest descent as well as conjugate gradient require $\nu$ steps to find the optimal tensor parameters and then their computational cost reads $\mathcal{O}(\nu d\chi^{3}) + \mathcal{O}(\nu d^{2}\chi^{2})$.
With steepest descent $\nu$ can get very large for the convergence of each each tensor update to good tensor parameters.
With conjugate gradient we make use of the fact that $-\Delta$ is positive definite -- and symmetric -- and then we can always find the optimal tensor parameters in each tensor update for $\nu = d\chi^{2}$.

The performance of the tensor train decomposition, i.e.\ matrix product states, in the context of the Poisson equation has been discussed in reference~\cite{OsDo12}, however, without MGR.
Here we want to identify the advantages of our MGR method.
MGR starts from the exact solution on a coarse grid -- i.e.\ the exact solution of the coarse grained finite difference problem -- and then successively refines the grid and finds solutions on the refined grids, until the solution for the desired number of grid points $N = 2^{L}$ is found.
In this article, our MGR method always starts from $L = 3$ and tensors with random entries, and we choose $\chi \geq d$ and perform alternating minimization until convergence to the exact solution.
We then repeatedly implement prolongation and minimization operations, as described in section~\ref{subsec:tps}, until the desired value of $L$ has been reached.
We compare our MGR method to the usual procedure which starts directly at the desired final value of $L$, uses tensors with random entries, and then performs alternating minimization.
We define that the alternating minimization procedure has converged when the relative change of the cost function is smaller than a given convergence precision.
In all our numerical experiments, we observe that, for a given convergence precision, MGR converges faster than the usual procedure and to lower final values of the cost function.
E.g.\ for convergence precision $10^{-4}$ with MGR, when the conjugate gradient method is used for the tensor update, and we consider a problem in one spatial dimension with a random polynomial $|g\rangle_{h}$, we find that the number of required alternating minimization sweeps is $\approx$ $10$ for $L = 6$, $12$ for $L = 7$, $21$ for $L = 8$, $29$ for $L = 9$, and $39$ for $L = 10$.
For the same convergence precision without MGR and the same problem, we find that the number of required alternating minimization sweeps is $\approx$ $29$ for $L = 6$, $57$ for $L = 7$, and $> 100$ for $L = 8$.
Although the quantitative details of our analysis (e.g.\ the required number of sweeps and final value of the cost function) are problem-specific (and depend on the spatial dimensionality, $|g\rangle_{h}$, and so on), qualitatively we always observe the same: MGR improves the convergence and final result.
We find that the advantages of MGR are most visible when steepest descent or conjugate gradient are used for the tensor update.
In these cases, it is not possible for us to converge with the usual procedure when $L$ is larger than $\approx 8$.
In contrast, MGR converges well for all values of $L$ that we have considered, namely up to $L = 20$.

\section{MGR for the ground state computation of the nonlinear Schr\"{o}dinger equation}\label{sec:nlse}

In this section MGR is used to develop a new algorithm for the ground state computation of the nonlinear Schr\"{o}dinger equation.
This equation is defined via a Hamiltonian operator that, after discretization, takes on the form
\begin{eqnarray}\label{eq:NLSE}
H(|f\rangle) & = & -\frac{1}{2}\Delta + V + g|f|^{2} .
\end{eqnarray}
$H$ depends on the function $|f\rangle$ that it acts upon and thus represents a nonlinear function.
$\Delta$ is the Laplace operator, $V$ represents an external potential and $g|f|^{2}$ the interactions.
Here we use this nonlinear Schr\"{o}dinger equation to describe Bose-Einstein condensates and references~\cite{DaGiPiSt99, PiSt03} provide details on the physical meaning of the individual parts of $H$ (note that equation~\ref{eq:NLSE} is also used in other contexts, e.g.\ nonlinear optics~\cite{Sco05}).
We are interested in the ground state, which is the eigenstate $|f\rangle$ corresponding to the smallest possible eigenvalue $E$ in the equation $H(|f\rangle) |f\rangle = E |f\rangle$.

\begin{figure}
\centering
\includegraphics[width=68.687mm]{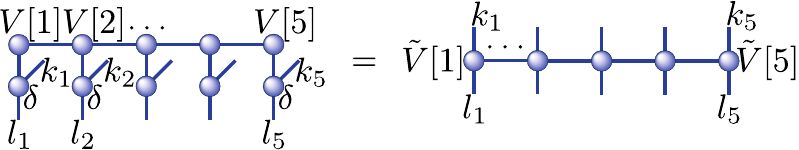}
\caption{\label{fig:6}
The matrix product state $V[1] V[2] \ldots V[5]$ is transformed into the matrix product operator $\tilde{V}[1] \tilde{V}[2] \ldots \tilde{V}[5]$ that is diagonal and has the values of the matrix product state on its diagonal.
This is achieved by contracting each individual tensor of the matrix product state with the $\delta$ tensor as shown.
The $\delta$ tensor has the entries $\delta_{\alpha, \beta, \gamma} = 1 \forall \alpha = \beta = \gamma$ and all other entries are zero.
}
\end{figure}

\begin{figure}
\centering
\includegraphics[width=71.114mm]{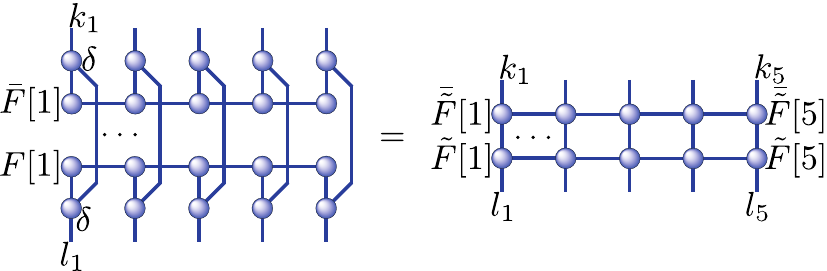}
\caption{\label{fig:7}
The matrix product state $F[1] F[2] \ldots F[5]$ is transformed into a tensor product operator that is diagonal and has the values of $|f|^{2}$ on its diagonal.
Here $\bar{F}[1] \bar{F}[2] \ldots \bar{F}[5]$ is the complex conjugate of $F[1] F[2] \ldots F[5]$ and is constructed from the individual tensors $\bar{F}[\ell]$ that contain the complex conjugates of the entries of $F[\ell]$.
The $\delta$ tensor is the same as in figure~\ref{fig:6}.
We contract each individual tensor $F[\ell]$ and $\bar{F}[\ell]$ with a $\delta$ tensor to obtain $\tilde{F}[\ell]$ and $\bar{\tilde{F}}[\ell]$, respectively, as shown.
}
\end{figure}

The individual parts of $H$ in equation~\ref{eq:NLSE} can be written as tensor product operators.
We already discussed the Laplace operator in section~\ref{subsec:pois}.
We assume a simple form of the potential $V$ which, most of the time, will be a polynomial that can be constructed as explained in section~\ref{subsec:pois}.
This construction gives a matrix product state for $V$.
For equation~\ref{eq:NLSE} we need to transform this matrix product state into a matrix product operator that has the values of the matrix product state on its diagonal.
That is achieved with the help of a simple $\delta$ tensor which has the entries $\delta_{\alpha, \beta, \gamma} = 1 \forall \alpha = \beta = \gamma$ and all other entries are zero -- note that this tensor is also known as the copy tensor~\cite{AlClFoJa11, DeBiJaCl12}.
Then the construction of $V$ as a matrix product operator is straightforward as can be seen in figure~\ref{fig:6}.
The interaction term $|f|^{2}$ needs to be constructed as a matrix product operator that has the modulus squared of the entries of $|f\rangle$ on its diagonal.
Using the $\delta$ tensor this construction is also straightforward as shown in figure~\ref{fig:7}.

\subsection{Quantifying the accuracy of MGR}\label{subsec:prec}

A standard algorithm for the ground state computation of the nonlinear Schr\"{o}dinger equation is imaginary time evolution~\cite{DaGiPiSt99, EdBu95}.
In this section we use this algorithm to identify quantities for measuring the accuracy of MGR.
Imaginary time evolution is based on the fact that $e^{-t H} |f\rangle$ converges to the eigenstate of $H$ corresponding to its smallest eigenvalue, i.e.\ the ground state, after long enough time $t$, if the initial state $|f\rangle$ is not orthogonal to that eigenstate.
Note that it is important to normalize $|f\rangle$ such that $\langle f|f \rangle = 1$ is true during the propagation, as otherwise the norm of $|f\rangle$ can increase or decrease very quickly during imaginary time evolution.
We split the total time $t$ into $N_{\tau}$ small time steps $\tau$, i.e.\ $t = N_{\tau} \tau$, such that $e^{-t H} |f\rangle = (e^{-\tau H})^{N_{\tau}} |f\rangle$ can be solved by applying the operator $e^{-\tau H}$ to the initial state $|f\rangle$ for $N_{\tau}$ times.
We use the approximation $e^{-\tau H} \approx \mathds{1} - \tau H$ with an approximation error $\mathcal{O}(\tau^{2})$ for each time step $\tau$.
In imaginary time evolution the propagation for the initial time steps does not need to be accurate because we are only interested in an accurate final state of the evolution.
So we start the evolution with a large value of $\tau$ for the initial steps and then successively decrease $\tau$ during the evolution such that the final steps are much more accurate than the initial ones.
More precisely, we start the propagation with a large value of $\tau = 0.5$ (which is large for all $H$, i.e.\ $g$ and $V$, considered here), converge the evolved state $|f\rangle$ by propagating it for large enough $N_{\tau}$ time steps, then decrease $\tau$ by a factor $1/2$, converge $|f\rangle$, and so on, until all the results that we are interested in have converged.
In our tensor product formalism, we find the evolved state $|\tilde{f}\rangle$ after each time step from the state $|f\rangle$ before the time step by minimizing the cost function $c(|\tilde{f}\rangle) = |||\tilde{f}\rangle - (\mathds{1} - \tau H)|f\rangle||^{2}$ where $H$ is defined in equation~\ref{eq:NLSE}.
We minimize this cost function by means of the alternating minimization procedure of reference~\cite{VePoCi04}.
Then we normalize the evolved state such that $\langle\tilde{f}|\tilde{f}\rangle = 1$ to avoid numerical problems.
Thus for each time step, our time evolution algorithm has a computational cost that scales like $\mathcal{O}(Ld^{2}\chi^{5})$.
Note that the dominant contribution to the computational cost comes from the contraction of $|f|^{2}$.

Here we only want to analyse different quantities for measuring the accuracy of MGR and so we use a particularly simple external potential $V$ in all numerical experiments of this section.
Throughout this section, in equation~\ref{eq:NLSE}, $V$ is the box potential: i.e.\ $V$ is $0$ everywhere inside the interval $[0, 1)$ and infinitely large outside this interval.

\begin{figure}
\centering
\includegraphics[width=58.630mm]{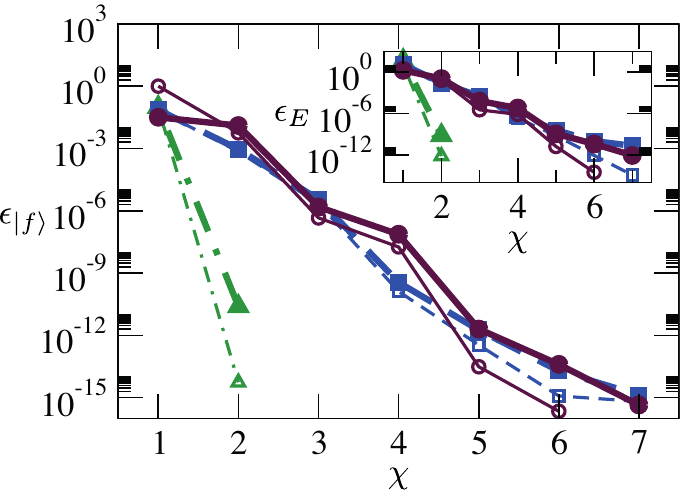}
\caption{\label{fig:8}
Infidelity $\epsilon_{|f\rangle}(\chi) := 1-|\langle f^{\chi}|f^{\mathrm{exact}}\rangle|$ (main) and relative energy error $\epsilon_{E}(\chi) := |E^{\chi}-E^{\mathrm{exact}}|/|E^{\mathrm{exact}}|$ (inset) as functions of the bond dimension $\chi$:
We consider $g = 0$ (dash-dotted, triangles), $g = 10$ (dashed, squares), and $g = 100$ (solid, circles), for $L = 6$ (thin lines, open symbols), and $L = 8$ (thick lines, filled symbols).
We observe that each error decreases exponentially with $\chi$, for all values of $g$ and $L$.
We also observe that, for a given value of $g$, the errors are almost the same for both sizes $L = 6$ and $8$.
For $g = 0$ the numerically exact solution is obtained with $\chi = 2$.
$\chi = 4$ to $7$ are required to reach machine accuracy for $g = 10$ and $100$.
}
\end{figure}

When the exact solution $|f^{\mathrm{exact}}\rangle$ is known, then we can compute the infidelity $\epsilon_{|f\rangle}(\chi) := 1-|\langle f^{\chi}|f^{\mathrm{exact}}\rangle|$ and the relative energy error $\epsilon_{E}(\chi) := |E^{\chi}-E^{\mathrm{exact}}|/|E^{\mathrm{exact}}|$.
Here, we compute the exact reference solution $|f^{\mathrm{exact}}\rangle$ by means of imaginary time evolution of a vector of dimension $N = 2^{L}$ which is initialized with random entries.
The errors $\epsilon_{|f\rangle}$ and $\epsilon_{E}$ are shown in figure~\ref{fig:8} for several values of $g$ and $L$.
We observe that all errors decrease exponentially as a function of $\chi$.
Because of this numerical evidence we can thus assume an upper bound of each error $\epsilon(\chi) \leq c_{0} \exp(-c_{1}\chi)$ where $c_{0}$ and $c_{1}$ are positive and real.
This implies that, to achieve a certain error, i.e.\ accuracy, $\epsilon$, we need to choose $\chi$ as
\begin{eqnarray}\label{eq:exp}
\chi(\epsilon) & = & \left\lceil \left( \log_{e}(c_{0}) - \log_{e}(\epsilon) \right)/c_{1} \right\rceil
\end{eqnarray}
where $\lceil$ and $\rceil$ denote the ceiling function and we assume that $\epsilon < 1$ such that $-\log_{e}(\epsilon) > 0$.
To gain a more intuitive understanding of equation~\ref{eq:exp}, we rewrite $-\log_{e}(\epsilon) = -log_{10}(\epsilon)/log_{10}(e)$, because we can understand $-log_{10}(\epsilon)$ as the number of digits that are correct for accuracy $\epsilon$.
Then equation~\ref{eq:exp} reads $\chi(\epsilon) = \lceil \tilde{c}_{0} - \tilde{c}_{1} \log_{10}(\epsilon) \rceil$ where $\tilde{c}_{0} = \log_{e}(c_{0})/c_{1}$ and $\tilde{c}_{1} = 1/(c_{1} \log_{10}(e))$.
We conclude that, if the error $\epsilon$ decreases exponentially as a function of $\chi$, then $\chi$ is a linear function of $-\log_{10}(\epsilon)$, i.e.\ $\chi$ grows linearly with the number of correct digits in the solution.
Equivalently, in this case, if we set the target accuracy $\epsilon$, then the required $\chi$ to achieve this accuracy is a linear function of $-\log_{10}(\epsilon)$.
Furthermore we can see in figure~\ref{fig:8} that the errors do not depend significantly on $L$.
Although we only compare $L = 6$ and $8$ (because we can only compute the exact solutions $|f^{\mathrm{exact}}\rangle$ for these relatively small values of $L$), we can clearly expect similar error functions for larger values of $L$ from the results shown in figure~\ref{fig:8}.
The fact that $\epsilon$ is independent of $L$ implies that, for a given value of $\chi$ we obtain the same accuracy for all possible values of $L$, i.e.\ number of grid points $N = 2^{L}$.
We conclude that tensor product states are an efficient ansatz for the problems considered here.

\begin{figure}
\centering
\includegraphics[width=61.843mm]{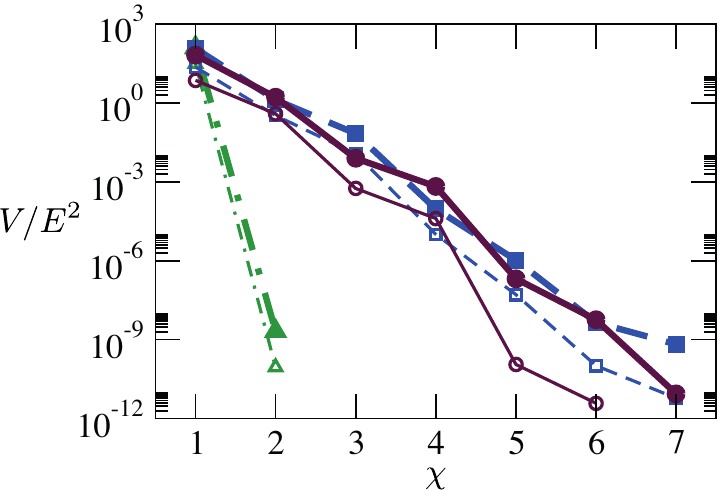}
\caption{\label{fig:9}
Normalized variance $V/E^{2}(\chi) := \langle f^{\chi}|H^{2}|f^{\chi}\rangle/\langle f^{\chi}|H|f^{\chi}\rangle^{2} - 1$ as a function of the bond dimension $\chi$:
We consider the same systems as in figure~\ref{fig:8}.
We observe that the normalized variance qualitatively behaves the same as the quantities shown in figure~\ref{fig:8}.
This suggests that the quantity shown here can be used to identify convergence of our method when the exact solution $|f^{\mathrm{exact}}\rangle$ is not known.
}
\end{figure}

For large values of $L$, i.e.\ number of grid points $N = 2^{L}$, we do not know $|f^{\mathrm{exact}}\rangle$ and therefore cannot compute $\epsilon_{|f\rangle}$ or $\epsilon_{E}$.
So we need to come up with alternative quantities to estimate the accuracy of tensor product state approximations.
The normalized variance $V/E^{2}(\chi) := \langle f^{\chi}|H^{2}|f^{\chi}\rangle/\langle f^{\chi}|H|f^{\chi}\rangle^{2} - 1$ vanishes for eigenstates of $H$.
In reference~\cite{VaHaCoVe16} such a variance was used to study ground state convergence with tensor product states.
Figure~\ref{fig:9} indicates that also in our case the normalized variance can be used to analyze convergence to the ground state.
However, it is important to keep in mind that this quantity measures convergence to an eigenstate which does not have to be the ground state.
Furthermore, in our case the computation of $\langle f^{\chi}|H^{2}|f^{\chi}\rangle$ has a high computational cost which is dominated by the scaling $\mathcal{O}(d^{2}\chi^{7})$ of the contraction for $\langle f^{\chi}||f|^{2} |f|^{2}|f^{\chi}\rangle$.
This high computational cost makes the computation of the normalized variance difficult for large values of $\chi$ and so we will not use this quantity in the remainder of this article.
Nevertheless this can be a useful quantity in the context of other (nonlinear) partial differential equations if it can be computed more efficiently there.

\begin{figure}
\centering
\includegraphics[width=56.488mm]{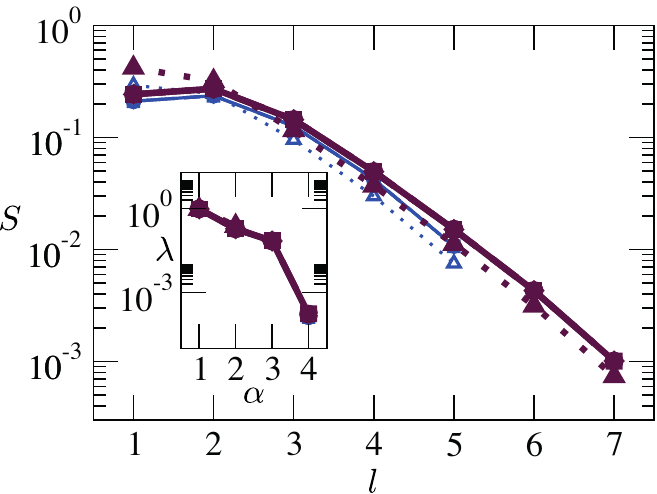}
\caption{\label{fig:10}
Entanglement entropy $S(l) := -\mathrm{tr}(\rho(l) \log_{2}(\rho(l)))$ where $\rho(l) := \mathrm{tr}_{1, 2, \ldots, l}(|f\rangle \langle f|)$ as a function of the level $l$ (main) and Schmidt coefficients $\lambda[l]_{\alpha}$ for level $l = 2$ (inset):
We consider $g = 100$, for $L = 6$ (thin lines, open symbols) and $8$ (thick lines, filled symbols), and $\chi = 2$ (dotted, triangles), $3$ (dash-dotted, diamonds), $4$ (dashed, squares), and $\chi_{\mathrm{max}}$ (solid, circles), where $\chi_{\mathrm{max}} = 8$ for $L = 6$ and $\chi_{\mathrm{max}} = 16$ for $L = 8$.
In the inset, we show the Schmidt coefficients for level $2$ because the exact solutions have the largest entanglement entropy for level $2$.
We observe that, for each value of $L$, all curves for $\chi \ge 3$ lie on top of each other.
We also observe that, for each value of $\chi$, our results for $L = 6$ and $8$ are almost the same.
We conclude that the problems considered here are only weakly entangled and can thus be described using low values of $\chi$ that are almost independent of the multigrid size $L$.
Therefore, our method converges rapidly with increasing bond dimension $\chi$, almost independently of $L$.
}
\end{figure}

The quantities that we use later to analyse the accuracy of MGR follow from a transformation of our matrix product state of equation~\ref{eq:mps} into its canonical form~\cite{Vi03, PeVeWoCi07, VeCiLaRiWo05}
\begin{eqnarray}\label{eq:canonicalmps}
|f\rangle & = & \sum_{l_{1}, l_{2}, \ldots, l_{L}} \sum_{\alpha_{1}, \alpha_{2}, \ldots, \alpha_{L-1}} \left( \Gamma[1]_{\alpha_{1}}^{l_{1}} \lambda[1]_{\alpha_{1}} \Gamma[2]_{\alpha_{1}, \alpha_{2}}^{l_{2}} \lambda[2]_{\alpha_{2}} \ldots \Gamma[L]_{\alpha_{L-1}}^{l_{L}} \right) |l_{1}\rangle |l_{2}\rangle \ldots |l_{L}\rangle
\end{eqnarray}
where each matrix $\lambda[l]$ is diagonal with nonnegative entries that correspond to the Schmidt coefficients for the bipartition of the state between $1$, $2$, \ldots, $l$ and $l+1$, $l+2$, \ldots, $L$.
A general indicator for the convergence of a state $|f^{\chi}\rangle$ with increasing $\chi$ is the convergence of Schmidt coefficients $\lambda[l]_{\alpha}$ and von Neumann entropies $S(l) := -\sum_{\alpha} (\lambda^{2}[l]_{\alpha} \log_{2}(\lambda^{2}[l]_{\alpha}))$ for all bipartitions of the state.
Notice that $S(l)$ is also known as the entanglement entropy $S(l) := -\mathrm{tr}(\rho(l) \log_{2}(\rho(l)))$ where $\rho(l) := \mathrm{tr}_{1, 2, \ldots, l}(|f\rangle \langle f|)$.
Figure~\ref{fig:10} shows example entanglement entropies and Schmidt coefficients for some of our problems.
With the help of the squared Schmidt coefficients $\lambda^{2}[l]_{\alpha}$, we can estimate the error~\cite{Os11, VeCi06}
\begin{eqnarray}\label{eq:error}
|||f^{\mathrm{exact}}\rangle - |f^{\chi}\rangle||^{2} & \leq & 2 \sum_{l=1}^{L-1} \sum_{\alpha_{l}=\chi+1}^{\min(d^{l}, d^{L-l})} \lambda^{2}[l]_{\alpha_{l}} .
\end{eqnarray}
Although we do not know the Schmidt coefficients $\lambda[l]_{\alpha}$ for $\alpha > \chi$, we do know them for $\alpha = 1$ to $\chi$ and can extrapolate their behavior for $\alpha = \chi+1$ to $\min(d^{l}, d^{L-l})$.
We can convince ourselves that our extrapolation is correct by increasing $\chi$ and checking that the new Schmidt coefficients lie on our extrapolated curve.
We have thoroughly investigated this procedure of extrapolating and checking the behavior of Schmidt coefficients, and in all cases our extrapolations were very accurate.
Therefore quickly decreasing Schmidt coefficients are an indicator for good accuracy of a matrix product state approximation.

Moreover, for all the problems considered here, we observe that the Schmidt coefficients for all bipartitions decrease exponentially as a function of $\chi$.
We thus assume that real and positive numbers $c_{0}$ and $c_{1}$ exist such that $\lambda[l]_{\alpha} \leq c_{0} \exp(-c_{1}\alpha)$ $\forall l$.
Then the error of equation~\ref{eq:error} has the upper bound
\begin{eqnarray*}
|||f^{\mathrm{exact}}\rangle - |f^{\chi}\rangle||^{2} & < & 2c_{0}^{2}(L-1) \sum_{\alpha=\chi+1}^{\infty} \exp(-2c_{1}\alpha)\\
& = & 2c_{0}^{2}(L-1) \exp(-2c_{1}(\chi+1)) \sum_{\alpha=0}^{\infty} \exp(-2c_{1}\alpha)\\
& = & 2c_{0}^{2}(L-1) \exp(-2c_{1}(\chi+1)) / (1-\exp(-2c_{1}))\\
& = & 2c_{0}^{2}(L-1) \exp(-2c_{1}\chi) / (\exp(2c_{1})-1)\\
& = & \tilde{c}_{0} (L-1) \exp(-\tilde{c}_{1}\chi)
\end{eqnarray*}
where $\tilde{c}_{0} = 2c_{0}^{2}/(\exp(2c_{1})-1)$ and $\tilde{c}_{1} = 2c_{1}$.
We conclude that if all Schmidt coefficients decrease exponentially as a function of $\chi$, then also the error defined in equation~\ref{eq:error} decreases exponentially as a function of $\chi$.
Using equation~\ref{eq:exp}, this implies that $\chi$ depends linearly on $-\log(\epsilon)$ -- where $\epsilon$ now denotes the error defined in equation~\ref{eq:error} -- and thus the required $\chi$ depends linearly on the number of desired correct digits.

Looking at all Schmidt coefficients is more informative than looking at just the von Neumann entropies.
Quick convergence of von Neumann entropies with increasing $\chi$ is a necessary but not sufficient criterion for good accuracy of a matrix product state approximation~\cite{ScWoVeCi08a}.

\subsection{Direct energy minimization via nonlinear tensor updates}\label{subsec:algo}

Imaginary time evolution is a reliable but inefficient algorithm for ground state computation, because the ground state is obtained in an indirect way via propagation in imaginary time for typically many time steps.
Direct minimization of the energy is more efficient as it aims at the ground state as directly as possible.
For the nonlinear Schr\"{o}dinger equation with the Hamiltonian of equation~\ref{eq:NLSE}, the energy is $E(|f\rangle) := \langle f|( -\frac{1}{2}\Delta + V + g|f|^{2})|f\rangle$ for normalized functions $|f\rangle$, i.e.\ that fulfill $\langle f|f\rangle = 1$.
The ground state minimizes $\langle f|( -\frac{1}{2}\Delta + V + \frac{g}{2}|f|^{2})|f\rangle$ under the constraint $\langle f|f\rangle = 1$ (notice that this ground state is equivalent to the normalized eigenstate of the Hamiltonian in equation~\ref{eq:NLSE} corresponding to the smallest eigenvalue)~\cite{LiYn98}.

We investigated several different methods for solving this constrained minimization problem from reference~\cite{GrNaSo09}.
The method that worked best substitutes the constrained minimization problem by the unconstrained minimization of a penalty function $P := \langle f|( -\frac{1}{2}\Delta + V + \frac{g}{2}|f|^{2})|f\rangle + \eta(\langle f|f\rangle-1)^{2}$ where $\eta$ is a new parameter.
In the limit $\eta \to \infty$ the minimum of this unconstrained minimization problem coincides with the minimum of the original constrained minimization problem.
Notice that a penalty function is a special cost function that results from replacing a constrained minimization problem by an unconstrained minimization problem and that depends on a penalty parameter in such a way that the desired solution is obtained when this penalty parameter is infinite.
To compute the ground state, we therefore need to minimize $P$ for successively growing values of $\eta$ until the energy $E(|f\rangle)$ has converged to its smallest value and $\langle f|f\rangle = 1$ is fulfilled well enough.
For each value of $\eta$, we minimize $P$ via an alternating minimization procedure that sweeps over the tensors and for each tensor minimizes $P$ via the variational parameters of this tensor only, i.e.\ keeping the parameters of all other tensors fixed.
Here each tensor update is a nonlinear optimization problem.
In the following we compare steepest descent, Fletcher-Reeves conjugate gradient, Polak-Ribi\`{e}re conjugate gradient, and Newton method when solving the nonlinear equations.
Additionally we compare the direct minimization of the penalty function with MGR and without MGR.

\begin{figure}
\centering
\includegraphics[width=57.942mm]{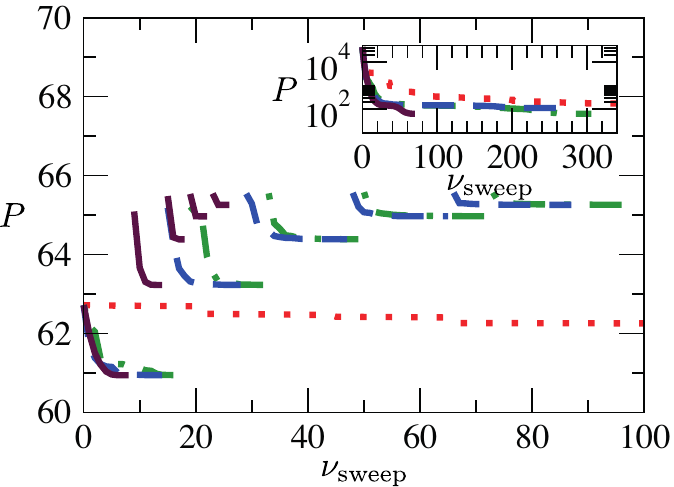}
\caption{\label{fig:11}
Penalty function $P := \langle f|( -\frac{1}{2}\Delta + V + \frac{g}{2}|f|^{2})|f\rangle + \eta(\langle f|f\rangle-1)^{2}$ as a function of the number of sweeps $\nu_{\mathrm{sweep}}$ -- where during each sweep we perform one tensor update of each tensor in $|f\rangle$ going from left to right and then we perform one tensor update of each tensor in $|f\rangle$ going from right to left -- for our MGR method (main) and without our MGR method (inset).
We consider the box potential $V$, with $g = 100$, and set $\eta = 10^{8}$, $L = 8$, and $\chi = 4$.
We compare four different methods for the nonlinear tensor update:
Each tensor update consists of $\nu_{\mathrm{update}} = 4$ steepest descent steps (dotted), $\nu_{\mathrm{update}} = 4$ Fletcher-Reeves conjugate gradient steps (dash-dotted), $\nu_{\mathrm{update}} = 4$ Polak-Ribi\`{e}re conjugate gradient steps (dashed), and $\nu_{\mathrm{update}} = 1$ Newton method step (solid).
Main (our MGR method):
We start from the exact solution for $L = 3$ (from imaginary time evolution), prolong it to $L = 4$, minimize the penalty function for $L = 4$ until convergence, prolong the solution to $L = 5$, minimize the penalty function for $L = 5$ until convergence, prolong the solution to $L = 6$, and so on.
For each $L$ the first plotted value of $P$ is computed directly after the prolongation.
E.g.\ for the Newton method (solid), the first continuous line (from $\nu_{\mathrm{sweep}} = 0$ to $\approx 10$) corresponds to $L = 4$, the second continuous line (from $\nu_{\mathrm{sweep}} \approx 10$ to $\approx 15$) corresponds to $L = 5$, the third continuous line (from $\nu_{\mathrm{sweep}} \approx 15$ to $\approx 20$) corresponds to $L = 6$, the fourth continuous line (from $\nu_{\mathrm{sweep}} \approx 20$ to $\approx 25$) corresponds to $L = 7$, and the fifth continuous line (from $\nu_{\mathrm{sweep}} \approx 25$ to $\approx 30$) corresponds to $L = 8$.
Inset:
We start from a random state for $L = 8$ and minimize the penalty function until convergence.
We define convergence when the relative change of the penalty function from one sweep to the next is smaller than $10^{-6}$.
}
\end{figure}

Our results are shown in figure~\ref{fig:11}.
We observe that our MGR method converges reliably as all four methods of the nonlinear tensor update considered here converge to the same final result of $P$ within reasonable relative precisions.
We take the final result of the Newton method as the reference value as it is the lowest value of $P$.
Then we obtain a relative precision of $2 \cdot 10^{-3}$ for steepest descent, $4 \cdot 10^{-5}$ for Fletcher-Reeves conjugate gradient, and $2 \cdot 10^{-5}$ for Polak-Ribi\`{e}re conjugate gradient.
The steepest descent method requires the largest number of sweeps (more than $2000$) for convergence and converges to the least accurate final result (i.e.\ the highest value of $P$).
Both conjugate gradient methods require similar numbers of sweeps for convergence (around $90$) and converge to similar final results.
The Polak-Ribi\`{e}re is a little bit faster and more precise than the Fletcher-Reeves conjugate gradient method.
The Newton method requires the smallest number of sweeps for convergence and gives the most accurate final result (i.e.\ the lowest value of $P$).
Notice that each time $L$ is increased by $1$ the penalty function converges to a larger value.
This is not an indication of the result getting worse but arises because of the normalization of $|f\rangle$ changing with $L$.

Each time $L$ is increased by $1$, the Newton method requires approximately $5$ sweeps until convergence while each conjugate gradient method requires approximately $20$ sweeps until convergence.
Firstly, this suggests that our MGR method requires a number of sweeps that grows linearly with $L$: $\nu_{\mathrm{sweep}} \propto L$ (notice that the accuracy of our finite discretization method grows exponentially with $L$).
Secondly, this suggests that the Newton method converges the fastest:
The computational cost of each tensor update is $\mathcal{O}(\chi^{6})$ in the Newton method and $\mathcal{O}(\chi^{5})$ in both conjugate gradient methods, and we performed $\nu_{\mathrm{update}} = 4$ ($= \chi$) updates per tensor for both conjugate gradient methods and $\nu_{\mathrm{update}} = 1$ update per tensor for the Newton method.
We can read off from the plot that, still, each conjugate gradient method requires approximately $4$ times more sweeps than the Newton method:
Thus, in total, each conjugate gradient method needs approximately $16$ $(= \chi^{2})$ times more tensor updates than the Newton method.

We can see in the inset of figure~\ref{fig:11} that without our MGR method no reliable convergence occurs.
The four methods considered here for the nonlinear tensor update do not converge to the same values of $P$.
Both the steepest descent and the Polak-Ribi\`{e}re conjugate gradient method converge to relatively high values of $P$, i.e.\ the methods possibly get stuck in high-lying local minima of $P$.
The Newton method achieves the lowest value of $P$ and this value coincides with the one obtained from our MGR method within a relative precision of $5 \cdot 10^{-11}$.
Without our MGR method significantly more sweeps are required for convergence than with our MGR method.

\begin{figure}
\centering
\includegraphics[width=67.323mm]{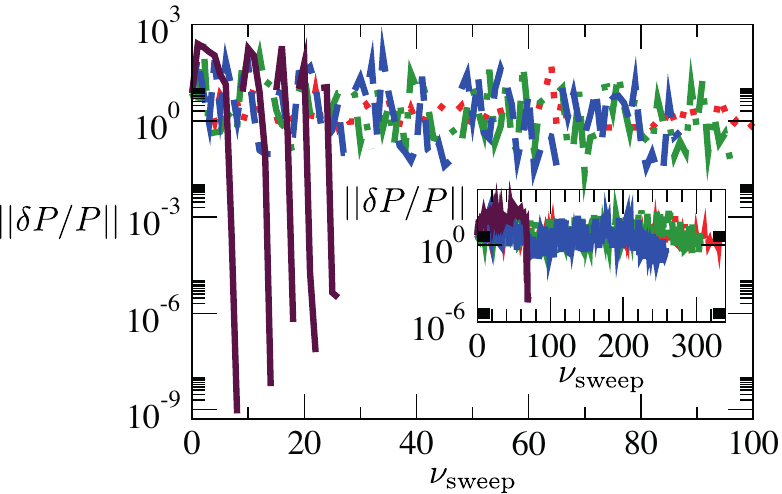}
\caption{\label{fig:12}
Norm of the normalized gradient of the penalty function $||\delta P / P|| := \sqrt{\sum_{l=1}^{L} ||\partial P / \partial \vec{F}[l]||^{2}} / |P|$ as a function of the number of sweeps $\nu_{\mathrm{sweep}}$ for our MGR method (main) and without our MGR method (inset).
We consider the same systems as in figure~\ref{fig:11}.
Both with and without our MGR method, the Newton method converges to a small penalty function gradient, i.e.\ a proper extremum of the penalty function.
With our MGR method this convergence is more systematic, quicker, and finishes in a lower final value than without our MGR method.
All the other tensor update methods converge less systematically, slower, and to significantly larger penalty function gradients than the Newton method.
}
\end{figure}

\begin{figure}
\centering
\includegraphics[width=56.933mm]{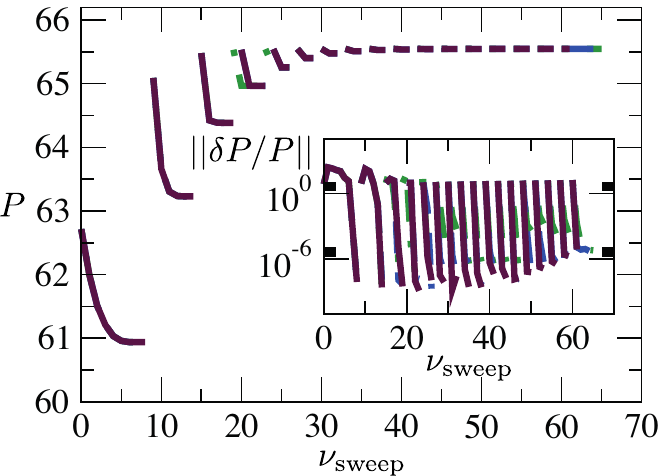}
\caption{\label{fig:13}
Penalty function $P$ (main) and norm of the normalized gradient of the penalty function $||\delta P / P||$ (inset) as a function of the number of sweeps $\nu_{\mathrm{sweep}}$ for our MGR method with $\nu_{\mathrm{update}} = 1$ Newton method step per tensor update.
We consider the box potential $V$, with $g = 100$, set $\eta = 10^{8}$, and $L = 20$, and compare $\chi = 4$ (dash-dotted), $8$ (dashed), and $16$ (solid).
For our MGR method we use the same procedure as in the main plot of figure~\ref{fig:11} with the same convergence precision $10^{-6}$.
We observe that our results for the different values of $\chi$ lie on top of each other for all values of $L \leq 20$.
This suggests that increasing $\chi$ further, i.e.\ setting $\chi > 16$, would have a negligible effect on all our results:
Therefore our MGR method has converged with $\chi = 16$.
We also observe that the convergence of our MGR method with increasing $\chi$ occurs for all values of $L \leq 20$ equally fast:
This suggests that the convergence with $\chi$ is independent of $L$.
Furthermore we read off from the plot that our MGR method requires a number of sweeps $\nu_{\mathrm{sweep}}$ to converge that grows linearly with $L$: $\nu_{\mathrm{sweep}} \propto L$.
}
\end{figure}

Convergence can be quantified via the norm of the normalized gradient of the penalty function $||\delta P / P|| := \sqrt{\sum_{l=1}^{L} ||\partial P / \partial \vec{F}[l]||^{2}} / |P|$.
This quantity is shown in shown in figure~\ref{fig:12}.
We draw the same conclusion from figure~\ref{fig:12} as we did from figure~\ref{fig:11}:
The combination that works best is MGR with the Newton method for the nonlinear tensor update.
This combination is applied to a finer grid, namely $L = 20$, in figure~\ref{fig:13}.
Figure~\ref{fig:13} illustrates a fact that is true for all problems considered in this article:
For a given convergence precision the number of sweeps required by our MGR method scales linearly with $L$.
Therefore our MGR method features an exponential speedup compared to the original MG method, namely from a computational cost scaling like $\mathcal{O}(N)$ to $\mathcal{O}(L) = \mathcal{O}(\log(N))$.

\subsection{Performance}\label{subsec:perf}

\begin{figure}
\centering
\includegraphics[width=58.775mm]{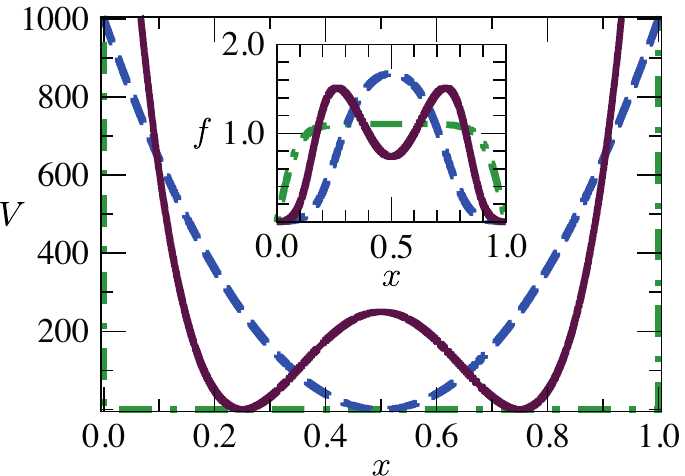}
\caption{\label{fig:14}
Main:
Potentials considered here: box $V_{\mathrm{b}}(x)$ (dash-dotted), well $V_{\mathrm{w}}(x)=4000(x-0.5)^{2}$ (dashed), and double-well $V_{\mathrm{dw}}(x)=250-8000(x-0.5)^{2}+64000(x-0.5)^{4}$ (solid).
Inset:
Ground state wave functions $f(x)$ of the one-dimensional nonlinear Schr\"{o}dinger equation with $g = 100$ for the potentials of the main plot.
We minimized the penalty function $P$ with $\eta = 10^{8}$, using our MGR method with $\nu_{\mathrm{update}} = 1$ Newton method step per tensor update and convergence precision $10^{-6}$, i.e.\ the same procedure as in figure~\ref{fig:13}.
These results correspond to $L = 20$ and $\chi = 24$.
We systematically compared the results shown here with the corresponding results for $L = 12$ and $20$ using $\chi = 12$, $16$, and $20$, and we found no significant differences between all these results.
Therefore we assume that choosing $L > 20$ or $\chi > 24$ will have a negligible effect on the results shown here.
Thus our MGR method has converged with increasing $L$ and $\chi$ to the true continuous ground state wave function.
}
\end{figure}

\begin{figure}
\centering
\includegraphics[width=53.137mm]{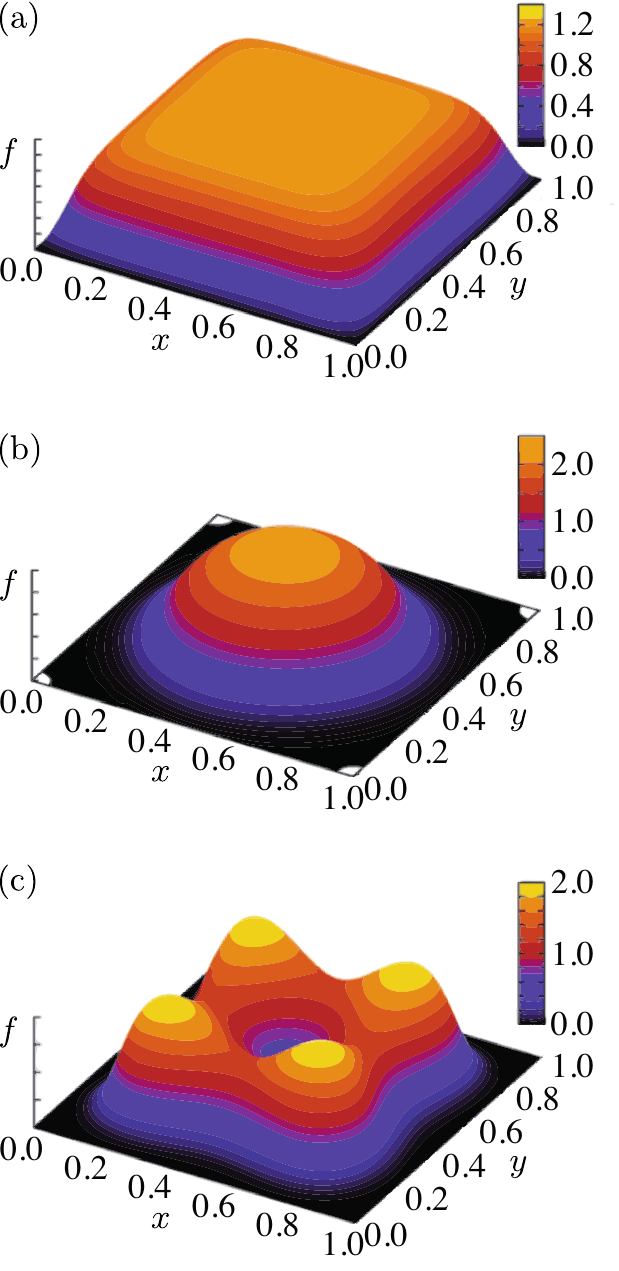}
\caption{\label{fig:15}
Ground state wave functions $f(x, y)$ of the two-dimensional Schr\"{o}dinger equation with $g = 100$ for the following potentials: box (a), well (b), and double-well (c) (as defined in figure~\ref{fig:14}).
We minimized the penalty function $P$ with $\eta = 10^{8}$, using our MGR method with $\nu_{\mathrm{update}} = 1$ Newton method step per tensor update and convergence precision $10^{-6}$, i.e.\ the same procedure as in figure~\ref{fig:13}.
These results correspond to $L = 20$ and $\chi = 24$ restricted to $L = 10$.
We systematically compared the results shown here with the corresponding results for $L = 12$ and $20$ using $\chi = 12$, $16$, and $20$, and we found no significant differences between all these results.
Therefore we assume that choosing $L > 20$ or $\chi > 24$ will have a negligible effect on the results shown here.
Thus our MGR method has converged with increasing $L$ and $\chi$ to the true continuous ground state wave function.
}
\end{figure}

We now analyse the performance of our MGR method by computing ground states of the nonlinear Schr\"{o}dinger equation in one, two, and three spatial dimensions.
We consider three different external potentials: a box, a well, and a double-well.
These external potentials are defined and shown for one spatial dimension in figure~\ref{fig:14}.
Notice that the behavior of the solution for a double-well potential depends on the ratio between barrier height and zero-point energy of the unperturbed single-well oscillator: if the barrier is too low, then we just have a perturbed (anharmonic) single-well problem, whereas if the barrier is too high, then the problem reduces to two well-isolated single-well problems.
Our double-well potential is at the cross-over between these two regimes.
Our external potentials for higher spatial dimensions are constructed from the ones for one spatial dimension by taking the same potential in each dimension.
E.g.\ the double-well potential reads $V_{\mathrm{dw}}(x) + V_{\mathrm{dw}}(y)$ in two and $V_{\mathrm{dw}}(x) + V_{\mathrm{dw}}(y) + V_{\mathrm{dw}}(z)$ in three spatial dimensions, where $V_{\mathrm{dw}}$ is defined in the caption of figure~\ref{fig:14}.
The corresponding matrix product operators are constructed in the same way that we used for the higher-dimensional Laplace operator in section~\ref{subsec:pois}.
E.g.\ for the double-well potential we use $V_{\mathrm{dw}} \otimes \mathds{1} + \mathds{1} \otimes V_{\mathrm{dw}}$ in two and $V_{\mathrm{dw}} \otimes \mathds{1} \otimes \mathds{1} + \mathds{1} \otimes V_{\mathrm{dw}} \otimes \mathds{1} + \mathds{1} \otimes \mathds{1} \otimes V_{\mathrm{dw}}$ in three spatial dimensions, where $V_{\mathrm{dw}}$ is the matrix product operator for one spatial dimension.
We choose these external potentials because the corresponding nonlinear Schr\"{o}dinger equations describe Bose-Einstein condensates in laser potentials as they are currently realised in many experiments all around the world.
Figures~\ref{fig:14} and~\ref{fig:15} illustrate how our solutions look like for one and two spatial dimensions, respectively.

\begin{figure}
\centering
\includegraphics[width=56.800mm]{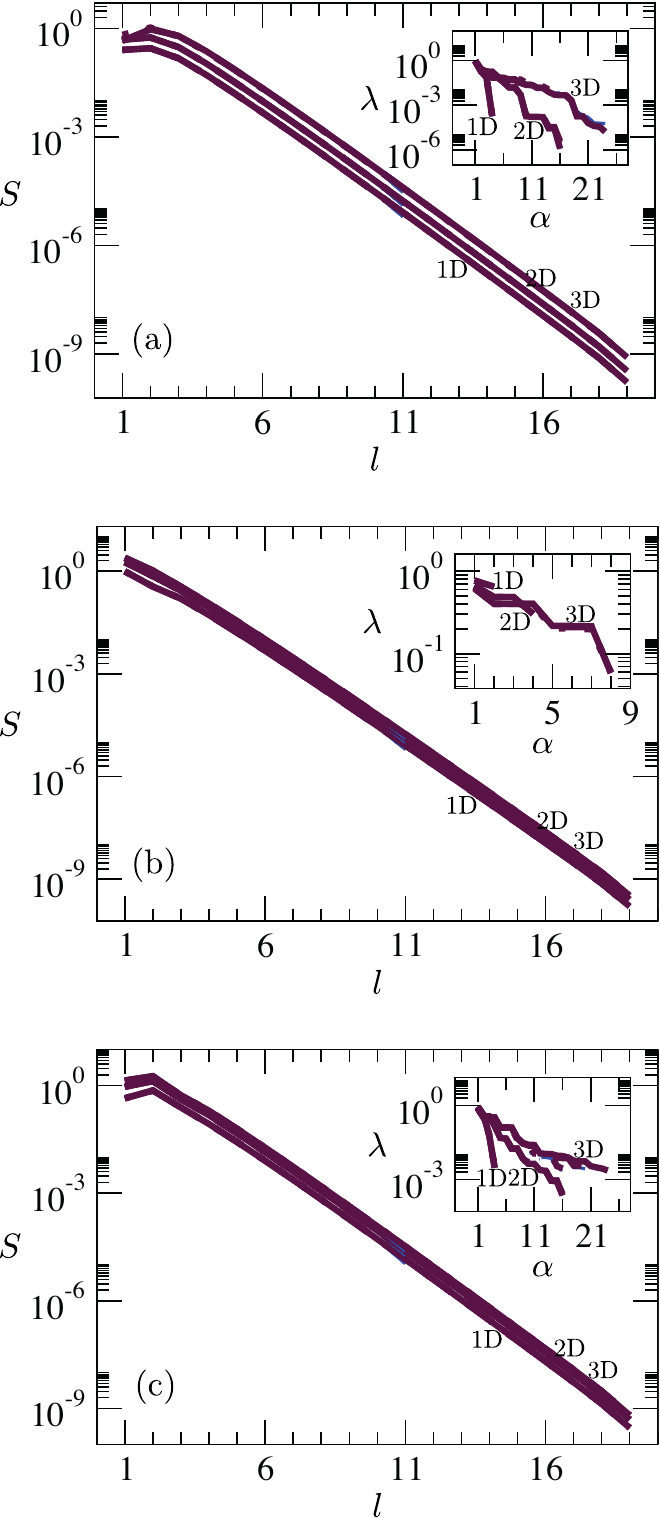}
\caption{\label{fig:16}
Entanglement entropy $S(l) := -\mathrm{tr}(\rho(l) \log_{2}(\rho(l)))$ where $\rho(l) := \mathrm{tr}_{1, 2, \ldots, l}(|f\rangle \langle f|)$ as a function of the level $l$ (main) and Schmidt coefficients $\lambda[l]_{\alpha}$ (inset) for level $l = 2$ (a), $1$ (b), and $2$ (c):
We consider the box $\hat{V}_{\mathrm{b}}$ (a), well $\hat{V}_{\mathrm{w}}$ (b), and double-well $\hat{V}_{\mathrm{dw}}$ (c) (as defined in figure~\ref{fig:14}), with $g = 100$ for $L = 12$ (thin lines) and $20$ (thick lines), and $\chi = 12$ (dotted), $16$ (dash-dotted), $20$ (dashed), and $24$ (solid).
In the inset, we show the Schmidt coefficients for the level with the largest entanglement entropy.
These results were obtained via minimization of the penalty function $P$ with $\eta = 10^{8}$, using our MGR method with $\nu_{\mathrm{update}} = 1$ Newton method step per tensor update and convergence precision $10^{-6}$, i.e.\ the same procedure as in figure~\ref{fig:13}.
We observe that in 1D and 2D all curves lie on top of each other.
In 3D the curves corresponding to $\chi > 16$ lie on top of each other and only the $\chi = 12$ and $16$ curves can be slightly off.
We conclude that the convergence of our MGR method with $\chi$ is independent of $L$ and that our MGR method has converged with $\chi = 24$, i.e.\ setting $\chi > 24$ will have a negligible effect on all our results.
}
\end{figure}

We investigate the convergence of our solution function $|f^{\chi}\rangle$ with increasing bond dimension $\chi$ and number of grid points $N$ in figure~\ref{fig:16}.
This allows us to give precise estimates for the ground state energies in continuous space, see table~\ref{tab:1}.
We observe in figure~\ref{fig:16} that the entanglement entropy grows when the spatial dimension increases.
This indicates that, to obtain a fixed accuracy for the solution, in higher spatial dimensions we need larger values of $\chi$.
Table~\ref{tab:1} confirms this:
For a fixed value of $\chi$, the accuracy decreases with increasing spatial dimensionality.
In this analysis we set the maximum $\chi$ to $24$ for all spatial dimensions.
However, we have converged results for $\chi = 64$ in one and $\chi = 48$ in two spatial dimensions.
For all problems considered here, we obtained machine precision of our solutions in one and two spatial dimensions.
Only in three spatial dimensions, where $d = 8$, it is difficult to obtain converged results for $\chi > 24$ because of limited computing power.
Nevertheless it is remarkable that such a small value of $\chi = 24$ suffices to describe $N = 8^{20} \approx 10^{18}$ grid points in three spatial dimensions and produces ground state energy accuracies that are better than one percent.

\begin{table}
\caption{\label{tab:1}
Ground state energies $E$ for the nonlinear Schr\"{o}dinger equation with $g = 100$, for the problems described in the text.
We minimized the penalty function $P$ with $\eta = 10^{8}$, using our MGR method with $\nu_{\mathrm{update}} = 1$ Newton method step per tensor update and convergence precision $10^{-6}$, i.e.\ the same procedure as in figure~\ref{fig:13}.
The number in brackets denotes the uncertainty of the last digit.
E.g.\ for 1D Box $E = 122.09942(7)$ means that the true result is in the interval $[122.09935, 122.09949]$.
We construct these numbers from the ground state energies for $L = 20$ and $\chi = 24$ (for 1D Box: $E = 122.09942$) and the corresponding ground state energies for $L = 19$ and $\chi = 20$ (for 1D Box: $E = 122.09935$).
We assume that the energy error decreases sufficiently fast with increasing number of grid points $N$ and bond dimension $\chi$.
Then the energy values provided here with error bars represent the true ground state energies for continuous space $N \to \infty$ and infinite bond dimension $\chi \to \infty$.
}
\centering
\begin{tabular}{p{3cm}p{3cm}p{3cm}p{3cm}}
\hline
 Dimensionality & Box          & Well         & Double-well  \\
\hline
 1D             & 122.09942(7) & 288.05273(3) & 264.67755(3) \\
 2D             & 145.0192(2)  & 515.2060(2)  & 444.6485(5) \\
 3D             & 197.0(3)     & 759.5(5)     & 640(1) \\
\hline
\end{tabular}
\end{table}

\begin{figure}
\centering
\includegraphics[width=75.560mm]{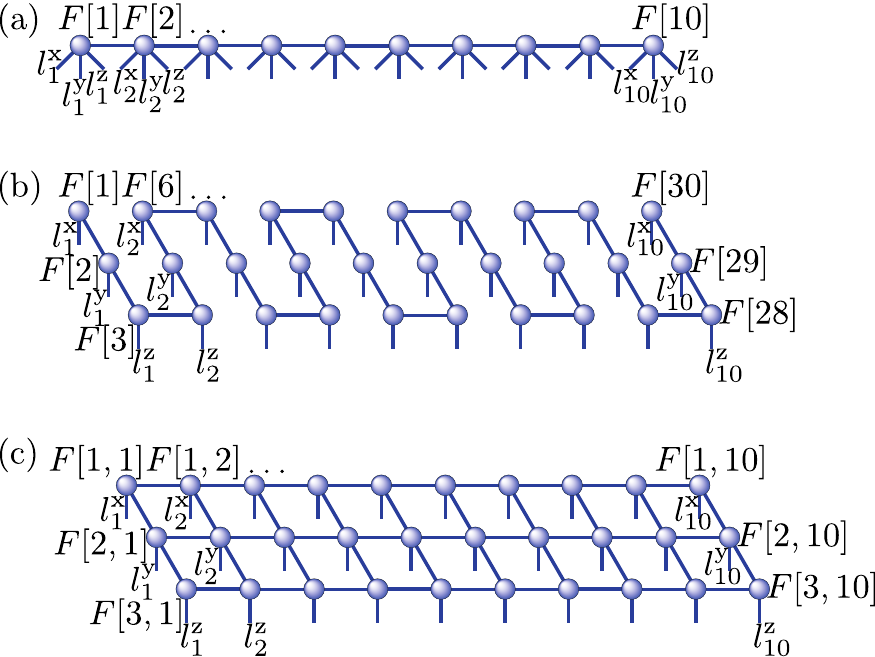}
\caption{\label{fig:17}
Different tensor product state ansatzes for three spatial dimensions.
(a) Matrix product state, as used in this article.
(b) Matrix product state, constructed as for quantum many-body systems in two spatial dimensions~\cite{StWh12}.
Compared to (a), (b) has the advantage that $d = 2$ -- and not $d = 8$ as in (a) -- and therefore has a lower computational cost.
(c) Projected entangled pair state~\cite{VeCi04, MuVeCi07}.
Compared to (a) and (b), (c) has the advantage that it can capture larger von Neumann entropies.
In particular, (c) can capture a larger increase of von Neumann entropies with the spatial dimension.
}
\end{figure}

Our results in figure~\ref{fig:16} and table~\ref{tab:1} indicate that for higher spatial dimensions a different tensor product state ansatz might be better suited.
Two alternative ansatzes are presented in figure~\ref{fig:17}.
A particularly promising ansatz is the finite projected entangled pair state~\cite{VeCi04, MuVeCi07} shown in figure~\ref{fig:17} (c).
This ansatz can capture much larger von Neumann entropies in higher spatial dimensions than matrix product states.

\section{Conclusions and outlook}

We have shown that MG methods for solving (nonlinear) partial differential equations generate functions in the form of tensor product states.
Therefore these methods can alternatively be formulated as variational algorithms over tensor product states.
This has the advantage that finer grids can be reached than with previous MG methods.
We have constructed our MGR method as a natural combination of MG and variational renormalization group concepts.
Our MGR method generalizes and extends the MG method.
Moreover, our MGR method has a computational cost that scales like $\mathcal{O}(\log(N))$ for $N$ grid points, whereas for the fastest previous MG methods the computational cost scales like $\mathcal{O}(N)$.
We have developed MGR algorithms for the linear Poisson and nonlinear Schr\"{o}dinger equation.
For all problems considered in this article, we have verified that the computational cost indeed scales like $\mathcal{O}(\log(N))$ and, thus, our MGR method is exponentially faster than previous MG methods.

In this article, we have only used matrix product states as our tensor product state ansatz for functions in any spatial dimension.
We have identified Schmidt coefficients and von Neumann entropies as useful quantifiers for the accuracy of a solution.
For our ground state solutions of the nonlinear Schr\"{o}dinger equation, we have often seen a significant increase of von Neumann entropies with the spatial dimension.
This can be better captured by another tensor product state ansatz called projected entangled pair states~\cite{VeCi04, MuVeCi07}.
It would be very interesting to apply the state-of-the-art finite projected entangled pair state algorithms of references~\cite{LuCiBa14a, LuCiBa14b} to partial differential equations.

In all our algorithms, we have found that the Newton method is more efficient and accurate for nonlinear tensor updates than steepest descent and conjugate gradient based approaches.
This might also be the case in other tensor product state algorithms.
For example, infinite projected entangled pair states~\cite{JoOrViVeCi08} are composed of repeating unit cells of tensors.
So far, most algorithms that use this ansatz to approximate ground states are based on imaginary time evolution.
Only recently algorithms have been proposed that rely on the more efficient direct energy minimization, see e.g.\ references~\cite{VaHaCoVe16, Co16, LiDoHaGuHe17}.
In this case, nonlinear tensor updates are required and we believe that the Newton method could accelerate the existing algorithms.

All problems that we have considered in this article have had relatively smooth solutions.
It would be interesting to analyze the performance of our MGR method in other examples like e.g.\ turbulence (see e.g.\ reference~\cite{NaGaSmHa16} for details on how turbulence can be observed in experiments described by nonlinear Schr\"{o}dinger equations).
The optimal tensor product state algorithm for dynamical problems is known~\cite{LuOsVa15, HaLuOsVaVe16}.
And so it would be straightforward to analyze the performance of our MGR method for such time-dependent problems.
However, there are many other interesting applications for MGR.
For example, it would be exciting to apply MGR to the systematic construction of new density functionals for density functional theory, using the methods developed in reference~\cite{LuFuApRuCiBa16}.
While the scenario of reference~\cite{LuFuApRuCiBa16} was limited to one spatial dimension and a relatively rough grid for the spatial coordinates as well as the density values, MGR can handle continuous functions in three spatial dimensions.
By combining MGR with the procedure presented in reference~\cite{LuFuApRuCiBa16}, we could construct a hierarchy of successively more precise exchange-correlation energy functionals for continuous density values in three spatial dimensions: an outstanding goal in density functional theory.
Another exciting future direction of research would be to address the natural question whether or not it is possible to apply our MGR method to the computation of excited states.

\section{Acknowledgements}

ML and DJ acknowledge funding from the NQIT (Networked Quantum Information Technologies) Hub of the UK National Quantum Technology Programme and from the EPSRC Tensor Network Theory Grant (EP/K038311/1).
All tensor product algorithms of this article were programmed using the Tensor Network Theory Library~\cite{AlClJa17}.

\bibliographystyle{model1-num-names}
\bibliography{refs}

\end{document}